# Measurements and TCAD Simulations of Bulk and Surface Radiation Damage Effects in Silicon Detectors

F. Moscatelli, P. Maccagnani, D. Passeri, G.M. Bilei, L. Servoli, A. Morozzi,
G.-F. Dalla Betta, R. Mendicino, M. Boscardin and N. Zorzi

*Abstract–* In this work we propose the application of a radiation damage model based on the introduction of deep level traps/recombination centers suitable for device level numerical simulation of radiation detectors at very high fluences (e.g. $1 \div 2 \times 10^{16}$ 1-MeV equivalent neutrons per square centimeter) combined with a surface damage model developed by using experimental parameters extracted from measurements from gamma irradiated p-type dedicated test structures.

## I. INTRODUCTION

IN the last few decades extensive simulations and experimental studies have been carried out to understand the mechanisms of radiation damage in solid-state silicon sensors for High-Energy Physics experiments [1-6]. During sensor operations at Large Hadron Collider (LHC), the high fluence of radiation introduces defects, both in the Si substrate causing bulk damage and at the Si-SiO$_2$ interface causing surface damage. Bulk damage impacts on the detector performance by introducing acceptor- and donor-type deep-level traps. This leads to higher leakage currents, change in the effective space charge concentration and lowering the charge collection efficiency due to charge trapping. Surface damage introduces oxide charges, in particular at the Si-SiO$_2$ interface, and interface traps, which strongly influence the break-down voltage, the inter-electrode isolation and capacitance, and might also impact the charge collection properties of silicon sensors.

In the past we developed a device-level model based on the introduction of three deep levels able to reproduce the radiation damage macroscopic effects up to fluences of the order of $10^{15}$ 1-MeV equivalent neutrons (n$_{eq}$)/cm$^2$[1]. The much higher fluences (at least a factor ten higher) expected at the High Luminosity LHC (HL-LHC), impose new challenges and the extension of the model to foresee the detector behavior at such fluences is not straightforward. New effects have to be taken into account such as avalanche multiplication and capture cross section dependencies on temperature and fluences, at the same time keeping the solid physically based approach of the modeling (e.g., by using no fitting parameters).

Moreover the voltage stability and the charge-collection properties of segmented silicon sensors are strongly influenced by the charge layers and potential distribution developing at the sensor surface, due to the radiation induced charge density variation within the oxide and passivation layers, and Si-SiO$_2$ interface trap states. The addition of these features to the model will preserve the generality of the approach, allowing its application to the optimization of different kind of detectors.

To better understand in a comprehensive framework these complex and articulated phenomena, measurements on test structures and sensors, as well as TCAD simulations related to bulk, surface and interface effects, have been carried out in this work. In particular, the properties of the SiO$_2$ layer and of the Si-SiO$_2$ interface, using MOS capacitors, MOSFETs and gate-controlled diodes (gated diodes) manufactured on high-resistivity p-type silicon substrate before and after irradiation with gamma rays at doses between 10 Mrad and 500 Mrad have been studied. In fact, while many studies have been carried out on <100> and <111> *n*-type silicon substrates [3, 4], to our knowledge surface effects on *p*-type silicon substrates have not been thoroughly explored, thus motivating the choice to focus this work on *p*-type silicon only.

## II. EXPERIMENTAL MEASUREMENTS

Test structures under study include gated-diodes, MOS capacitors and MOSFET fabricated at FBK (Trento, Italy) on 6", *p*-type SiSi Direct Wafer Bonded Wafers from ICEMOS Technology Ltd [7]. The active layer is a Float Zone, <100> wafer, with a nominal resistivity higher than 3 kΩ·cm. The fabrication technology is "*n*-in-*p*", with a p-spray layer implanted at the surface to isolate adjacent n$^+$ electrodes. Moreover test structures include MOS capacitors fabricated at Institute for Microelectronics and Microsystems (IMM) of the Italian National Research Council of Bologna (Italy) on 4", *p*-type <100> silicon wafer with a nominal resistivity of 1Ω·cm.

**Manuscript received.** This work was supported by the H2020 project AIDA-2020, GA no. 654168.

Francesco Moscatelli is with CNR - IMM Bologna, Via Gobetti 101, 40129 Bologna, Italy, e-mail: moscatelli@bo.imm.cnr.it (F. Moscatelli) and with INFN Perugia, via Pascoli, 06125 Perugia, Italy.

Piera Maccagnani is with CNR - IMM Bologna, Via Gobetti 101, 40129 Bologna, Italy

Daniele Passeri and Arianna Morozzi are with DI of University of Perugia and with INFN of Perugia, via G. Duranti 93, 06131 Perugia, Italy.

Gian Mario Bilei and Leonello Servoli are with INFN of Perugia, via Pascoli, 06125 Perugia, Italy.

Gian-Franco Dalla Betta and Roberto Mendicino are with DII of University of Trento and TIFPA-INFN, Via Sommarive 9,38123 Trento, Italy.

Maurizio Boscardin and Nicola Zorzi are with Fondazione Bruno Kessler (FBK) and TIFPA-INFN, Via Sommarive 18, 38123 Trento, Italy.

We considered 15 test structures manufactured by FBK of Trento and 6 test structures with 4 MOS capacitors for each die manufactured by IMM of Bologna.

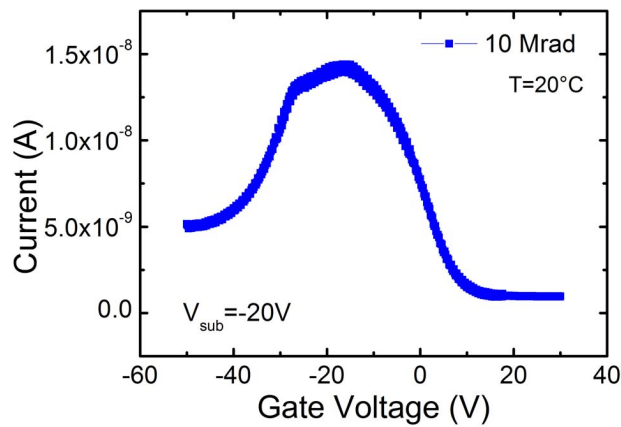

Fig. 1: Diode current as a function of gate voltage for an interdigitated gated diode after a gamma dose of 10 Mrad and an annealing at 80°C for 10 min.

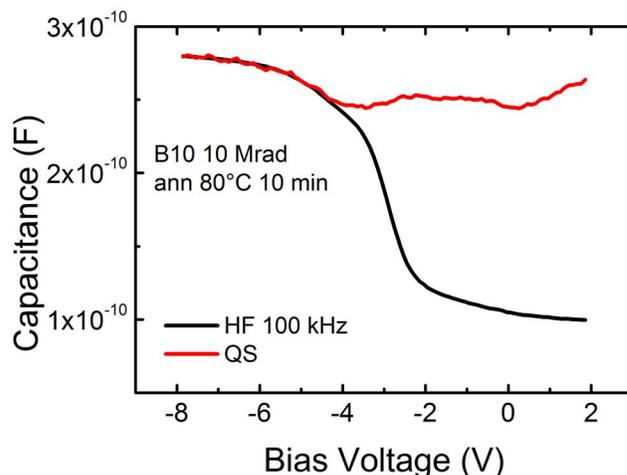

Fig. 2 High-frequency and quasi-static capacitance measured on a MOS capacitor after a gamma dose of 10 Mrad and an annealing at 80°C for 10 min.

Current-Voltage (I-V) and Capacitance-Voltage (C-V) measurement have been carried out at 20°C using a computer-controlled parametric system. This is based on a semiautomatic probe station (Micromanipulator P200A) with a HC-1000 thermal chuck system, and an Agilent B1500A semiconductor device analyzer. Using High-Frequency (HF) and Quasi-Static (QS) C–V characteristics and I-V measurements, the oxide-charge density ($N_{OX}$), the surface velocity ($s_0$) and the interface-trap density ($D_{IT}$) near the valence band have been determined before and after irradiation with gamma rays. From the diode current as a function of the gate voltage of a gated diode we calculated the surface velocity $s_0$ which is of the order of 0.5 cm/s before irradiation for all the measured test structures. The bias polarization of the substrate ($V_{SUB}$) is -2V. For the C-V characterization the HF measurements were carried out at 100 kHz with a small signal amplitude of 15 mV. The QS characteristics were measured with delay times of 0.7 sec using a voltage step of 100 mV. The interface state density was estimated by using the High-Low method [8].

Before irradiation the oxide charge density is of the order of $2\times10^{10}$ cm$^{-2}$, extrapolated from flat-band voltages of about -1 V. The interface trap density is of the order of $10^9$-$10^{10}$ cm$^{-2}$ eV$^{-1}$ in the range 0.2-0.6 eV from the valence band.

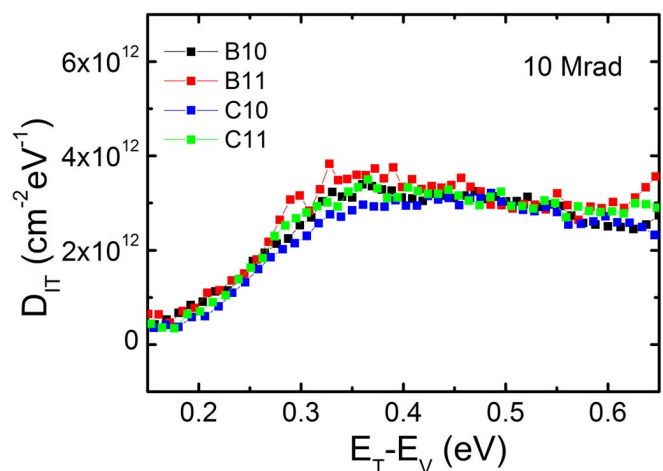

Fig. 3: Interface state density ($D_{it}$) near the valence band measured on four MOS capacitors after a gamma dose of 10 Mrad and an annealing at 80°C for 10 min.

Gamma-ray irradiations are being performed at the Gamma Irradiation Facility of Sandia National Laboratories (USA) in the range 10-500 Mrad. Measurements have been carried out after an annealing at 80°C for 10 min.

After irradiation from the diode current as a function of the gate voltage of a gated diode we calculated the surface velocity $s_0$ at $V_{SUB}$=-20 V, which is of the order of 500 cm/s (Fig. 1) . The surface velocity is similar in the range 10-500 Mrad for all the measured test structures. The integrated density of interface traps ($N_{IT}$) related to this surface velocity has been estimated as $7\times10^{11}$ cm$^{-2}$ following the procedure reported in [9].

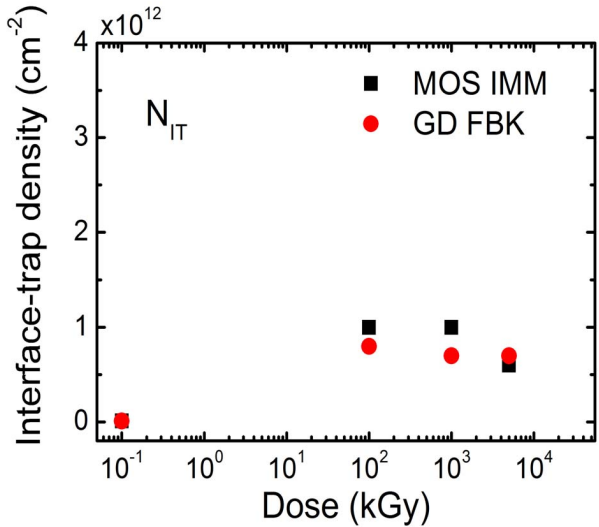

Fig. 4: The integrated density of interface traps ($N_{it}$) as a function of the gamma dose measured on MOS capacitors and gated diodes after an annealing at 80°C for 10 min.

From the measurements on the MOS capacitors after irradiations we extracted an oxide charge density ($N_{OX}$) of the order of $1\times10^{12}$ cm$^{-2}$, extrapolated from flat-band voltages of about -3.5 V (Fig. 2). As illustrated in Fig. 3 the interface trap density is of the order of $3\times10^{12}$ cm$^{-2}$ eV$^{-1}$ in the range 0.3-0.6 eV from the valence band. By Integrating the interface trap density in the range 0.2-0.6 eV we obtain an integrated density of interface traps which is of the order of $1\times10^{12}$ cm$^{-2}$ (Fig. 4).

Integrated density of interface traps and oxide charge density extracted from C–V characteristics and current-voltage (I-V) measurements are summarized in Table I. By performing a de-convolution of the findings illustrated in Fig. 3 it is possible to obtain two Gaussian curves picked at $E_T=E_V+0.35$ eV and at $E_T=E_V+0.56$ eV, with a peak concentration of $2.5\times10^{12}$ cm$^{-2}$ eV$^{-1}$ and $2\times10^{12}$ cm$^{-2}$ eV$^{-1}$, respectively (Fig. 5). These levels will be analyzed by simulations to verify which of them has an effect on the interstrip resistance.

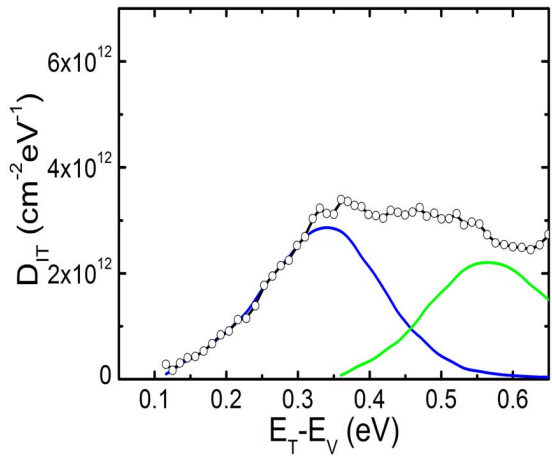

Fig. 5 Deconvolution of interface state density ($D_{it}$) near the valence band measured on MOS capacitors after a gamma dose of 10 Mrad and an annealing at 80°C for 10 min.

TABLE I
INTEGRATED DENSITY OF INTERFACE TRAPS ($N_{IT}$) AND OXIDE CHARGE DENSITY ($N_{OX}$) EXTRACTED FROM C–V CHARACTERISTICS AND CURRENT-VOLTAGE (I-V) MEASUREMENTS

|  | 10 Mrad | 100 Mrad | 500 Mrad |
|---|---|---|---|
| J (µA/cm$^{-2}$) GCD | 1.3 | 1.1 | 1.2 |
| $N_{IT}$ (cm$^{-2}$) GCD | $8\times10^{11}$ | $7\times10^{11}$ | $7\times10^{11}$ |
| $N_{OX}$ (cm$^{-2}$) MOS | $1.2\times10^{12}$ | $1.2\times10^{12}$ | $0.8\times10^{12}$ |
| $N_{IT}$ (cm$^{-2}$) MOS | $1\times10^{12}$ | $1\times10^{12}$ | $6\times10^{11}$ |
| $D_{IT}$ (cm$^{-2}$ eV$^{-1}$) MOS | 2.5-3x10$^{12}$ (0.3-0.6 eV) | 2.5-3x10$^{12}$ (0.3-0.6 eV) | 1.4-1.8 x10$^{12}$ (0.3-0.6 eV) |

## III. SIMULATIONS

### A. Bulk damage model

A comprehensive analysis of the variation of the effective doping concentration (Neff), the leakage current density and the charge collection efficiency (CCE) as a function of the fluence has been performed using the Synopsys Sentaurus TCAD device simulator. A simple one-dimensional structure was considered, consisting of a single diode 300 µm-thick with an high-resistivity p-type substrate ($3\times10^{12}$ cm$^{-3}$), without oxide allowing to focus on bulk effects. The trap levels used here are designed to model p-type Float Zone silicon after proton irradiation, and are based on work done by our group in the past [1]. The details of the traps are given in Table II and III. Each defect level is characterized by different parameters: the energy level (eV), the associated defect type, the cross section for electrons $\sigma_e$ and holes $\sigma_h$ (cm$^{-2}$) and the introduction rate $\eta$ (cm$^{-1}$), respectively [10].

TABLE II
THE RADIATION DAMAGE MODEL FOR P-TYPE
(UP TO $7\times10^{15}$ n/cm$^2$)

| Type | Energy (eV) | $\sigma_e$ (cm$^{-2}$) | $\sigma_h$ (cm$^{-2}$) | $\eta$ (cm$^{-1}$) |
|---|---|---|---|---|
| Acceptor | Ec-0.42 | $1\times10^{-15}$ | $1\times10^{-14}$ | 1.613 |
| Acceptor | Ec-0.46 | $7\times10^{-15}$ | $7\times10^{-14}$ | 0.9 |
| Donor | Ev+0.36 | $3.23\times10^{-13}$ | $3.23\times10^{-14}$ | 0.9 |

TABLE III
THE RADIATION DAMAGE MODEL FOR P-TYPE
(IN THE RANGE $7\times10^{15}$-2.2 x$10^{16}$ n/cm$^2$)

| Type | Energy (eV) | $\sigma_e$ (cm$^{-2}$) | $\sigma_h$ (cm$^{-2}$) | $\eta$ (cm$^{-1}$) |
|---|---|---|---|---|
| Acceptor | Ec-0.42 | $1\times10^{-15}$ | $1\times10^{-14}$ | 1.613 |
| Acceptor | Ec-0.46 | $3\times10^{-15}$ | $3\times10^{-14}$ | 0.9 |
| Donor | Ev+0.36 | $3.23\times10^{-13}$ | $3.23\times10^{-14}$ | 0.9 |

The simulated electrical characteristics of irradiated detectors have been compared with experimental measurements extracted from the literature, showing a very good agreement in terms of steady-state parameters (variation of depletion voltage, increase of the leakage current, electric field distribution) and in particular in terms of variation of the CCE as a function of the fluence.

In particular, in Figs. 6 and 7 the comparison between simulated and experimental charge collection in n-in-p strip detectors at 248 K and 500 V (Fig. 6) and 900 V (Fig.7) bias is reported. A very good agreement along the whole fluence range of operation at HL-LHC has been obtained .

Within this respect, the most effective level is the acceptor near the mid-gap ($E_C$-0.46 eV) . In particular, by changing the capture cross section for holes of this level (Table II and III) it is possible to reproduce the experimental data in terms of charge collection efficiency on p-type silicon pad detectors up to fluences of $2.2\times10^{16}$ $n_{eq}/cm^2$.

The avalanche generation effect which can be upset at such high fluences has been considered by comparing the different models available within Synopsys Sentaurus TCAD. The avalanche default model is the Van Overstraeten-De Man model [11]. We have tried to use as well the Lackner [12], the Okuto-Crowell [13] and the University of Bologna models[14]. The charge collection efficiency variation using different models lies in the range of a 3-4%. The important aspect to underline is the importance of the avalanche effect as illustrated in Fig. 8, in which the charge collection efficiency with and without avalanche effect is compared. Without the activation of the avalanche model the charge collection efficiency is poorly reproduced at very high fluences, showing the relevance of the multiplication effect on the charge generation.

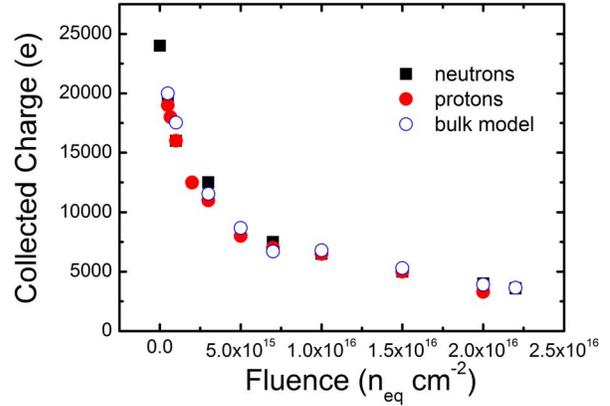

Fig.7 Comparison between simulated and experimental charge collection in n-in-p strip detectors at 248 K and 900 V bias. Measurements are taken by [15].

### B. Surface damage model

In order to develop a comprehensive bulk and surface damage model, the fixed oxide-charge density increase and additional interface trap states build-up with irradiation fluence have been considered as well. To evaluate the surface effect in term of strip isolation, a simple two strip structure with double p-stop structure (very low peak doping density of $5\times10^{15}$ cm$^{-3}$), 4 μm wide and separated by 6 μm has been considered. The bulk doping of sensor was $3\times10^{12}$ cm$^{-3}$.

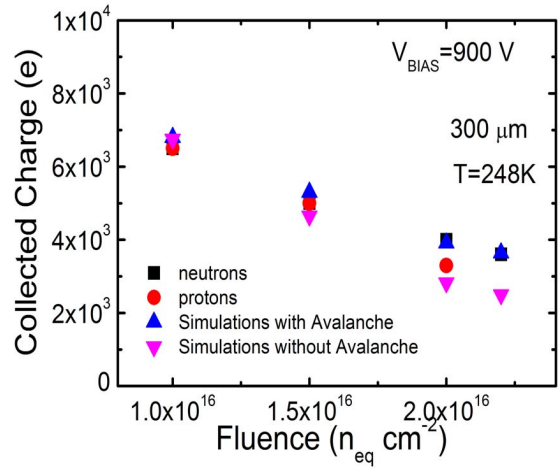

Fig.8 Comparison between simulated charge collection in n-in-p strip detectors at 248 K and 900 V bias with and without avalanche model.

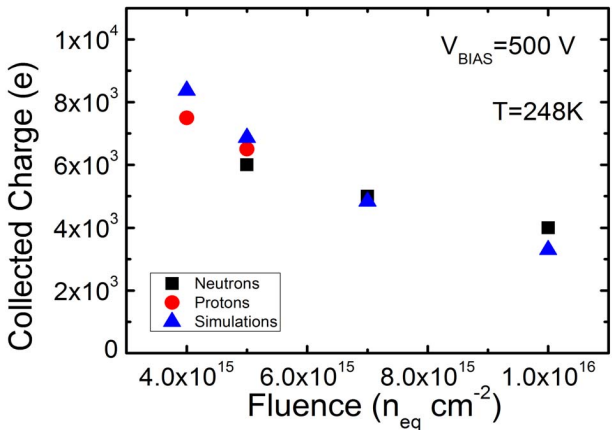

Fig. 6 Comparison between simulated and experimental charge collection in n-in-p strip detectors at 248 K and 500 V bias. Measurements are taken by [15].

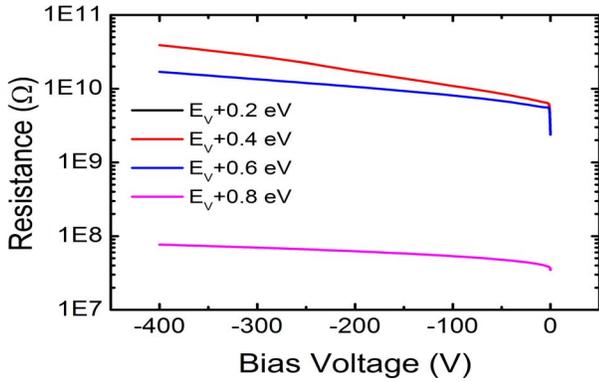

Fig.9 Simulated interstrip resistance as a function bias voltage at different levels of the donor interface traps. Simulations are obtained considering two acceptor interface traps at $E_T=E_C-0.4$ eV and at $E_T=E_C-0.6$ eV and one donor interface trap at different energy levels.

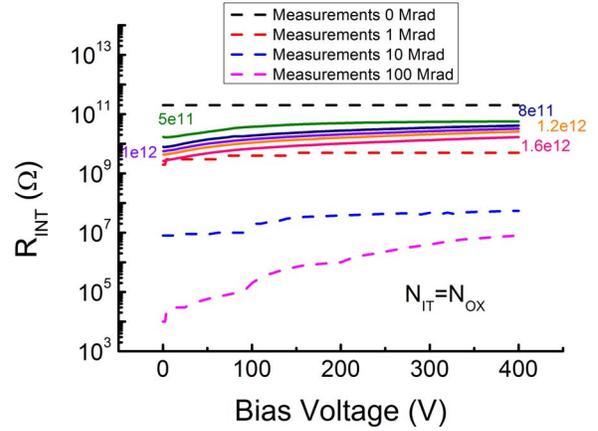

Fig.11 Measured [16] and simulated interstrip resistance as a function of the $V_{BIAS}$ at different x-ray doses. Simulations are obtained considering two acceptor interface traps at $E_T=E_C-0.4$ eV and at $E_T=E_C-0.6$ eV and one donor interface trap at Ev+0.6 eV. The coloured number are related to the oxide charge density ($N_{OX}$). In this case $N_{IT}= N_{OX}$

Fig. 10 shows the comparison of measured and simulated interstrip resistance as a function of bias voltage at different X-ray doses for the case $N_{IT}/N_{OX} =0.5$. Experimental data are taken from [16]. Different simulated values of resistance are obtained changing the $N_{OX}$. With this $N_{IT}/N_{OX}$ ratio is not possible to reproduce the interstrip resistance as a function of the dose because the isolation is poor for values of $N_{OX}$ of the order of $1 \times 10^{12}$ cm$^{-2}$. Fig 11 shows the same comparison for $N_{IT}/N_{OX} =1$. In this case the interstrip resistance is always very high even increasing very much $N_{OX}$ and the strips are always isolated.

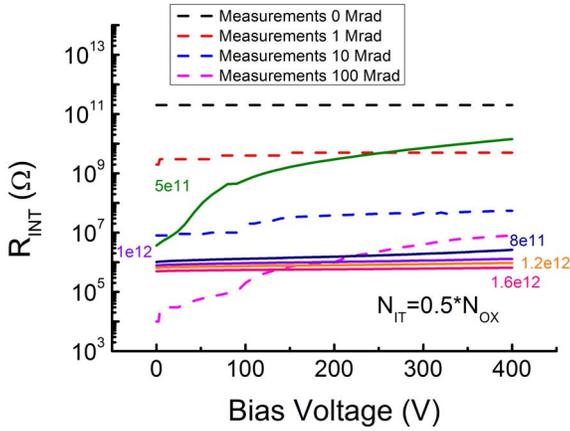

Fig.10 Measured [16] and simulated interstrip resistance as a function of the $V_{BIAS}$ at different x-ray doses. Simulations are obtained considering two acceptor interface traps at $E_T=E_C-0.4$ eV and at $E_T=E_C-0.6$ eV and one donor interface trap at Ev+0.6 eV. The coloured numbers are related to the oxide charge density ($N_{OX}$). In this case $N_{IT}=0.5*N_{OX}$

Considering the results obtained in [3, 4] we introduced two acceptor interface traps at $E_T=E_C-0.4$ eV and at $E_T=E_C-0.6$ eV. Once set these two interface traps the effect of the energy level of a third interface trap, a donor one, is analyzed for different doses, showing that only near the mid-gap the donor interface trap has an effect (Fig. 9). The level at $E_V+0.6$ eV from the valence band extracted from the measurements after gamma irradiation has been introduced in the model. The interface trap density ($N_{IT}$) measured in MOS capacitors is comparable to the oxide charge ($N_{OX}$. For a given acceptor concentration $N_{IT}$, 60% of acceptors traps are at $E_T=E_C-0.6$ eV, and 40% are traps at $E_T=E_C-0.4$ eV [3, 4]. We considered for the acceptor states and donor states the same $N_{IT}$ and analyzed different ratios $N_{IT}/N_{OX}$, in the range 0.5-1.

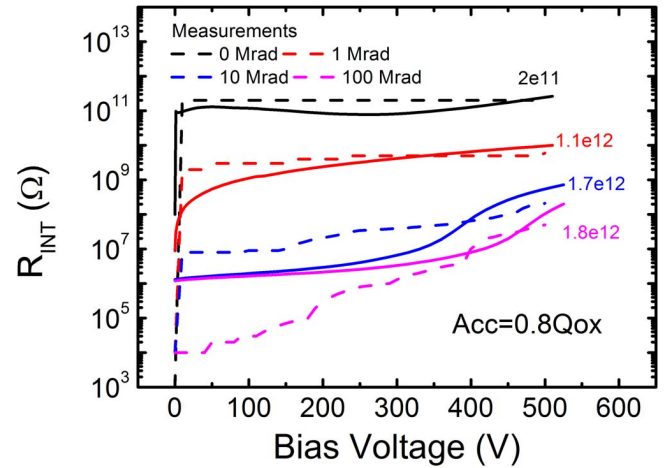

Fig.12 Measured [16] and simulated interstrip resistance as a function of the $V_{BIAS}$ at different x-ray doses. Simulations are obtained considering two acceptor interface traps at $E_T=E_C-0.4$ eV and at $E_T=E_C-0.6$ eV and one donor interface trap at Ev+0.6 eV. The coloured number are related to the oxide charge density ($N_{OX}$). In this case $N_{IT}= 0.8*N_{OX}$

Eventually, Fig. 12 shows the comparison of measured and simulated interstrip resistance as a function of bias voltage at different X-ray doses for the case $N_{IT}/N_{OX} =0.8$. Increasing the oxide charge and consequently the interface trap densities with the X-ray dose, it is possible to match the experimental data using the data described previously and summarized in Table IV.

TABLE IV
OXIDE CHARGE (NOX) AND INTERFACE TRAP DENSITY INTRODUCED IN THE SURFACE DAMAGE MODEL CONSIDERING LITERATURE DATA PUBLISHED IN [3]

| Interface Defect | Level | Concentration |
|---|---|---|
| Acceptor | $E_C$-0.4 eV | 40% of acceptor $N_{IT}$ ($N_{IT}$=0.8·$N_{OX}$) |
| Acceptor | $E_C$-0.6 eV | 60% of acceptor $N_{IT}$ ($N_{IT}$=0.8·$N_{OX}$) |
| Donor | $E_V$+0.6 eV | 100% of donor $N_{IT}$ ($N_{IT}$=0.8·$N_{OX}$) |

## IV. CONCLUSIONS

A new damage modelling scheme, suitable within commercial TCAD tools, featuring bulk and surface radiation damage effects has been proposed, aiming at extend prediction capabilities to high fluences HL-LHC radiation damage levels (e.g. fluences > $2.2\times10^{16}$ n/cm$^2$). The surface model has been developed introducing physical parameter values extracted from experimental measurements on gated diodes and MOS capacitors realized on p-type substrates after gamma irradiations in the range 10-500 Mrad. This model could be therefore used as a predictive tool for investigating sensor behavior at different fluences, temperatures and bias voltages for the optimization of 3D and planar silicon detectors foreseen at future HL-LHC High Energy Physics experiments.

ACKNOWLEDGMENT

The authors would like to thank Sally Seidel e Martin Hoeferkamp of University of New Mexico (USA) for gamma irradiations.